\documentclass{iaus}
\usepackage{graphicx}

\title[~~Interacing Luminous and Ultra-Luminous Infrared Galaxies ] 
{A Spitzer Study of Interacting Luminous and Ultra-Luminous Infrared
  Galaxies} \author[Jo\~ao Rodrigo S. Le\~ao \& Claus Leitherer]
{Jo\~ao Rodrigo S. Le\~ao$^1$ \and Claus Leitherer$^2$}

\affiliation{$^1$Universidade Federal do Rio Grande - FURG, \\ 
Instituto de Matem\'atica, Estat\'istica e F\'sica - IMEF\\
Caixa Postal 474, Campus Carreiros, Rio Grande, RS, Brazil\\ 
email: {\tt joaoleao@furg.br} \\[\affilskip]
$^2$Space Telescope Science Institute\\ 
3700 San Martin Drive, Baltimore, MD, USA\\
email: {\tt leitherer@stsci.edu}}

\pubyear{2011} 
\volume{284}  
\pagerange{1--12}
\jname{The Spectral Energy Distribution of Galaxies}
\editors{R.J. Tuffs \&  C.C.Popescu, eds.}
\begin{document}

\maketitle

\begin{abstract}

We conducted a Spitzer Space Telescope survey of 28 Luminous ($11 <
log (L_{IR}/L_\odot) < 12$, LIRGs) and Ultra-Luminous Infrared
Galaxies ($log (L_{IR}/L_\odot) > 12$, ULIRGs). Many of these galaxies
are found in pairs or associations and are powered by either nuclear
activity or star-formation (\cite[Sanders \& Mirabel
  1996]{Sanders_Mirabel_1996}). Our main goal is to understand the
relative importance of starbursts and AGNs in interacting systems. Is
the frequency of AGN and starbursts in these interacting galaxies
related to their luminosities? What is the importance of the merger
stage and the frequency of AGNs? We present our conclusions and
diagnostic diagrams based in the observed near infrared lines and
compare to studies based solely in optical data.

\end{abstract}

\firstsection 

\section{Introduction and Goals}

\begin{figure}[b]
\begin{center}
 \includegraphics[width=3.4in]{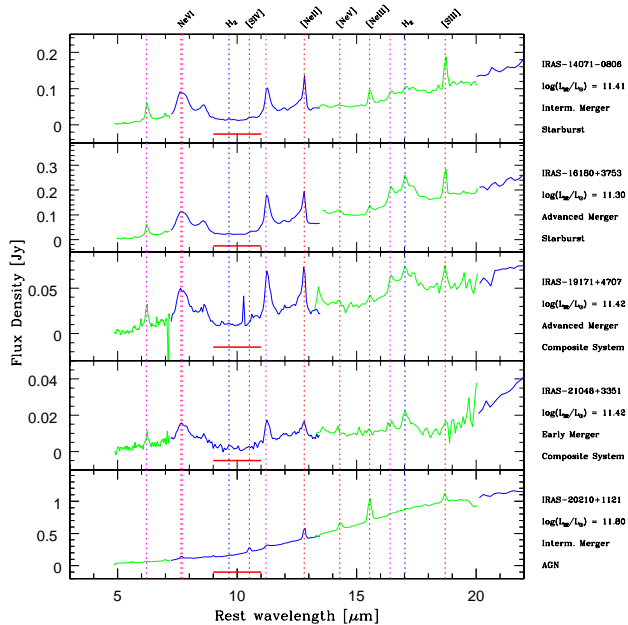}
 \caption{The IRS/SPitzer low­resolution spectra showing in detail the
   region between 4-22$\mu$m for the indicated galaxies,
   representative of the whole sample. Infrared luminosities, merger
   stages and activity class are also indicated. The main emission
   lines are indicated by vertical dotted lines: PAH lines, forbidden
   emission lines and molecular Hidrogen.  The horizontal bar
   indicates the extent of the silicate absorption around 10$\mu$m.}
\label{fig1}
\end{center}
\end{figure}

We present the results of a Spitzer Space Telescope survey of 28
Luminous and Ultra-Luminous Infrared Galaxies (\cite[Le\~ao \&
  Leitherer 2005]{Leao_Leitherer_2005}, \cite[Le\~ao \& Leitherer
  2008]{Leao_Leitherer_2008}). The observations were performed using
the IRS spectrograph in low-resolution to cover the 4–35 $\mu$m
wavelength range. We measured the flux and equivalent widths of the
several observed lines. We are particularly interested in emission
lines which characterize the presence of AGNs or starbursts as will be
discussed in section 2. Some of the observed SEDs are shown in figure
\ref{fig1}. In this study the LIRGs were chosen from the study of
Arribas et al. (2004) and images from the Nordic Optical telescope
(NOT) are available for these galaxies. The ULIRGs were chosen from
the study of Bushouse et al. (2002) and Hubble Space Telescope (HST)
images are available. Many of these galaxies are found in pairs or
associations (e.g., \cite[Bushouse et al. 2002]{Bushouse_etal_2002},
\cite[Arribas et al. 2004]{Arribas_etal_2004}) and are believed to be
powered by a combination of nuclear activity and star-formation (e.g.,
\cite[Genzel \& Cesarsky 2000]{Genzel_Cesarsky_2000}, \cite[Sanders \&
  Mirabel 1996]{Sanders_Mirabel_1996}).

The study of \cite[Sanders et al. (1988)]{Sanders_etal_1988} suggests
that the fraction of AGNs inreases with the luminosity of the host
galaxy.  In fact, the optical study of \cite[Kim, Veilleux \& Sanders
  (1998)]{Kim_Veilleux_Sanders_1998}, hereafter KVS98, indicates that
for LIRGs, AGNs are present in about 32\% of their galaxies, while for
ULIRGs, AGNs (Sey. 1, 2 and LINERs) dominate and are present in 70\%
of the systems. In many cases AGNs and starbursts coexist, making it
even harder to understand the relative importance of nuclear activity
and star-formation in LIRGs and ULIRGs.

The main goal of this project in to understand if diagnostics based in
the near infrared spectra yield the same results of KVS98. Do we find
the same relative frequency of starbursts and AGNs as a function of
luminosity? What is the importance of the merger stage and the
frequency of AGNs?

\section{Diagnostic Diagrams}

We used the observed near­IR spectra to determine the activity class
of our galaxies independently of their optical classification. The
presence of the PAH lines at 6.2, 7.7, and 11.3 $\mu$m are strong
indications for the presence of starbursts since the PAH molecules are
destroyed (or hidden) by nuclear activity (AGN). The presence of the
[Ne V] line at 14.3 and 24.3$\mu$m is a strong evidence for an AGN
since it takes 97 eV to ionize [NeIV] and this energy cannot come from
OB stars. The AGNs also exhibit a feature­free continuum below 5$\mu$m
that cannot be found in starbursts.

\begin{figure}[b]
\begin{center}
 \includegraphics[width=3.4in]{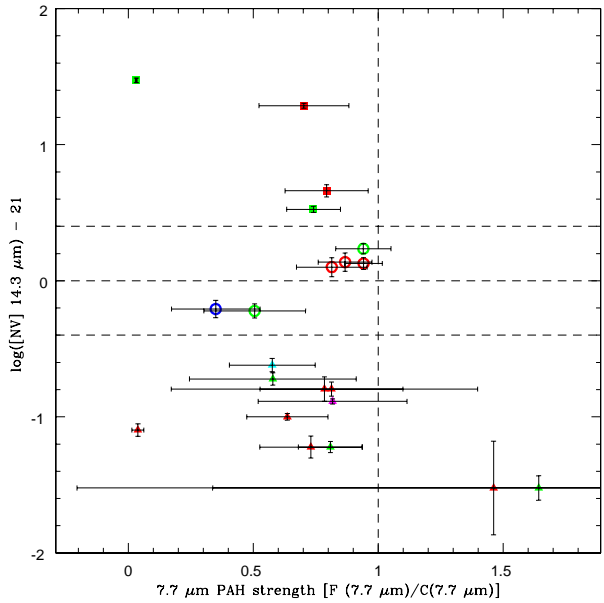}
 \caption{In this plot triangles denote starbursts, circles composite
   systems and squares AGNs. We see that AGNs and composite systems
   have stronger [N V] emission lines. Strong 7.7$\mu$m PAH lines are
   seen in AGN's, composites and starbursts, possibly indicating that
   star formation and nuclear activity might coexist in a single
   galaxy.} 
\label{fig2}
\end{center}
\end{figure}

In figure \ref{fig2} we investigate the strength of the 7.7 PAH
emission line vs. the [NeV] 14.3 $\mu$m emission line and find that
starbursts and AGN's are clearly separated. We designed other
diagnostic diagrams using the [OIV] at 25.9 $\mu$m and [NeII] at 12.8
$\mu$m. All these diagrams succesfully separate galaxies with AGN
activity from those powered by starbursts.

\section{Results}

Our main findings are:(i) Most AGNs are also advanced mergers. This is
in agreement with the \cite[Sanders et al. (1988)]{Sanders_etal_1988}
scenario to explain LIRGs and ULIRGs; (ii) We find that for $11 < log
(L_{IR}/L_\odot < 12$ (LIRGs) starbursts dominate with 75\% (25\%
AGNs), while for $log (L_{IR}/L_\odot) \geq 12$ (ULIRGs), AGNs and
composite systems dominate with 92\%. We thus find a strong evidence
for the so called luminosity dependence; (iii) Based on this small
sample we find no evidence that advanced mergers are preferentially
found in more luminous systems. In other words, we cannot claim that
the infrared luminosity is a precise tracer of merger activity. Maybe
a larger sample can shed some light on this issue. Many of these
galaxies show a deeep silicate absorption and this indicates a large
optical depth to the nucleus. We thus suspect that the AGN
contribution for many of the galaxies in the sample may be
larger. This would imply that AGNs remain hidden even at mid-IR
wavelengths.

\end{document}